\documentstyle[prl,aps,epsfig,amsmath,twoside]{revtex}
\begin{document}
\linespread{1.}
\twocolumn[\hsize\textwidth\columnwidth\hsize\csname@twocolumnfalse\endcsname 
\title{Crossing the dripline to $^{11}$N using elastic resonance
  scattering}
\author{\small
  K.~Markenroth$^{1}$,
  L.~Axelsson$^{1}$,
  S.~Baxter$^{2}$,
  M.~J.~G.~Borge$^{3}$, 
  C.~Donzaud$^{4}$,
  S.~Fayans$^{5}$, 
  H.~O.~U.~Fynbo$^{6}$,
  V.~Z.~Goldberg$^{5}$,
  S.~Gr\'{e}vy$^{10}$,
  D.~Guillemaud-Mueller$^{4}$,
  B.~Jonson$^{1}$,
  K.-M.~K\"{a}llman$^{7}$,
  S.~Leenhardt$^{4}$,
  M.~Lewitowicz$^{8}$, 
  T.~L\"{o}nnroth$^{7}$,
  P.~Manng\aa rd$^{7}$,
  I.~Martel$^{3}$, 
  A.~C.~Mueller$^{4}$,
  I.~Mukha$^{5,12}$,
  T.~Nilsson$^{6}$, 
  G.~Nyman$^{1}$,
  N.~A.~Orr$^{10}$,
  K.~Riisager$^{9}$, 
  G.~V.~Rogachev$^{5,14}$,
  M.-G.~Saint-Laurent$^{8}$,
  I.~N.~Serikov$^{5}$, 
  N.~B.~Shul'gina$^{1,5}$
  O.~Sorlin$^{4}$,
  M.~Steiner$^{2}$,
  O.~Tengblad$^{3}$, 
  M.~Thoennessen$^{2}$,
  E.~Tryggestad$^{2}$,
  W.~H.~Trzaska$^{13}$,
  F.~Wenander$^{1,6}$, 
  J.~S.~Winfield$^{10}$,
  R.~Wolski$^{11}$
\vspace{1mm}
}

\address
{\textit
{\small $^{1}$Experimentell fysik, Chalmers
 Tekniska H\"{o}gskola och G\"{o}teborgs Universitet,
    S-412 96 G\"{o}teborg, Sweden \\
\small  $^{2}$NSCL, Michigan State University, 
 East Lansing, MI--48824, USA \\
\small $^{3}$Instituto de Estructura de la Materia, CSIC, E--28006
         Madrid, Spain \\
 \small  $^{4}$Institut de Physique Nucl\'{e}aire, IN2P3--CNRS, F--91406
 Orsay Cedex, France \\
\small  $^{5}$Kurchatov Institute, Institute of General and
 Nuclear Physics, RU--123182 Moscow, Russia \\
  \small  $^{6}$EP Division, CERN, CH-1211 Geneva 23, Switzerland \\
 \small  $^{7}$Department of Physics, \AA bo Akademi, FIN--20500
 Turku, Finland \\
 \small  $^{8}$GANIL, BP 5027, F--14021 Caen Cedex, France \\
 \small  $^{9}$Institut for Fysik og Astronomi, Aarhus Universitet, 
 DK--8000 Aarhus C, Denmark \\
 \small  $^{10}$Laboratoire de Physique Corpusculaire, 
      F--14050 Caen Cedex, France \\
 \small $^{11}$Institute of Nuclear Physics, Cracow, Poland \\
  \small  $^{12}$Technische Universit\"{a}t, D--64289 Darmstadt, Germany \\
  \small  $^{13}$Department of Physics, University of Jyv\"{a}skyl\"{a}, 
 JYFL, FIN--40351  Jyv\"{a}skyl\"{a}, Finland \\
  \small $^{14}$Department of Physics, University of Notre Dame,
  IN-46556, USA
}
}
\maketitle

\vspace{1cm}
\begin{abstract}
The level structure of the unbound nucleus $^{11}$N has been 
studied by $^{10}$C+p elastic resonance scattering 
in inverse geometry with the LISE3 spectrometer at 
GANIL, using a $^{10}$C beam with an energy of 9.0~MeV/u. 
An additional measurement was done at the A1200 spectrometer at MSU.
The excitation function above the $^{10}$C+p threshold 
has been determined up to 5~MeV. A potential-model 
analysis revealed three resonance states 
at energies 1.27$^{+0.18}_{-0.05}$~MeV ($\Gamma$=1.44$\pm$0.2~MeV), 
2.01$^{+0.15}_{-0.05}$~MeV, ($\Gamma$=0.84$\pm$0.2~MeV) and 
3.75$\pm$0.05~MeV, ($\Gamma$=0.60$\pm$~0.05~MeV) with the
spin-parity assignments I$^{\pi}$=1/2$^{+}$, 1/2$^{-}$, 
5/2$^{+}$, respectively.
Hence, $^{11}$N is shown to have a ground state 
parity inversion completely analogous to its mirror partner,
$^{11}$Be. A narrow resonance 
in the excitation function at 4.33$\pm$0.05~MeV was also 
observed and assigned spin-parity 3/2$^{-}$.
 \\

\noindent
PACS number(s): 21.10.Hw, 21.10.Pc, 25.40.Ny, 27.20.+n
\end{abstract}

\vspace{12mm}
]

\section{Introduction}\label{sec:intro}
The exploration of exotic nuclei is one of the most 
intriguing and fastest expanding fields in modern nuclear physics. 
The research in this domain has introduced many new and 
unexpected phenomena of which a few examples are halo
systems, intruder states, soft excitation modes and 
rare $\beta$-delayed particle decays.
To comprehend the new features of the nuclear world that are
revealed as the drip-lines are approached, reliable and unambiguous
experimental data are needed. Presently available data for nuclei close
to the driplines mainly give ground-state properties as masses,
ground state $I^\pi$ and beta-decay half-lives. 
Also information on energies, widths and quantum numbers $I^{\pi}$ of
excited nuclear levels are vital for an understanding of the
exotic nuclei but are to a large extent limited to what can be
extracted from $\beta$ decays. Nuclear reactions 
can give additional information, in particular concerning 
unbound nuclear systems.
However, the exotic species are mainly produced in complicated reactions
between stable nuclei. These processes are normally 
far too complex to allow for spin-parity assignments of
the populated states, and  
hence are of limited use for spectroscopic investigations.
Instead of using complex reactions between stable nuclei, the
driplines can be approached in simple reactions involving radioactive nuclei.
An example is given in this paper where elastic resonance scattering of a
$^{10}$C beam on a hydrogen target was 
used to study the unbound nucleus $^{11}$N. 
With heavy ions as beam and light particles as target, the
technique employed here is performed in inverse geometry. The use of a
thick gas target instead of a solid target is another novel approach.  
This technique has been developed at the Kurchatov Institute~\cite{art90:52}
where it has been employed to study unbound cluster states with stable
beams~\cite{gol97:60}.
The perspectives of using radioactive beams in inverse kinematics
reactions to study exotic nuclei are discussed in~\cite{gol93:56} and
the method was used in~\cite{axe96:54}.
Resonance elastic scattering in inverse kinematics using radioactive beams 
and a solid target has been used at
Louvain-la-Neuve~\cite{cos94:50,ben92:321}.

This experiment is part of a large program for investigating 
the properties of halo states in nuclei~\cite{jon98:356}. 
A well studied halo nucleus is $^{11}$Be
where experiments have demonstrated that
the ground state halo mainly consists of an $1s_{1/2}$ neutron coupled to the
deformed $^{10}$Be core~\cite{ann93:304,nun96:596}, 
in contradiction to shell-model which predicts that the odd neutron 
should be in a $0p_{1/2}$ state. The $0p_{1/2}$ level is in reality    
the first excited state, while the ground state is a
$1s_{1/2}$ intruder level~\cite{mil83:28}. 
This discovery has been followed by numerous papers investigating the
inversion, e.~g.~references~\cite{tal60:4,sag93:309}.
The mirror nucleus of $^{11}$Be, $^{11}$N, should have a 1/2$^{+}$
ground state with the odd proton being mainly in the
$1s_{1/2}$ orbit, if the 
symmetry of mirror pairs holds. 
However, $^{11}$N is unbound with respect to proton
emission which means that all states are resonances that can be studied
in elastic scattering reactions.
The first experiment devoted to a study of the properties of the
low-energy structure of $^{11}$N 
used the three-nucleon transfer-reaction 
$^{14}$N($^{3}$He,$^{6}$He)$^{11}$N. The results
indicated a resonant state at 2.24~MeV~\cite{ben74:9} which 
was interpreted as the first excited 1/2$^{-}$ state rather than
the 1/2$^{+}$ ground state.

In this paper we present excitation functions at laboratory (lab.) angles of 
0$^{\circ}$ 
measured at GANIL~\cite{axe96:54} and MSU, and at 
12.5$^{\circ}$ with respect to the beam direction
measured at GANIL. 
A thorough analysis, using a potential model as well as a
simplified R-matrix treatment, gives unambiguous
determination of the quantum structure of the three lowest resonances in
the $^{10}$C+p system. 

\section{Elastic resonance scattering: method and formalism}\label{sec:expmeth}
The first description of elastic resonance scattering was given by Breit and
Wigner~\cite{bre36:49}, and it is now a theoretically well understood 
reaction mechanism~\cite{hod97,lau51:84b}. 
Traditionally, elastic resonance scattering experiments 
have been performed by bombarding a thin target with a
light ion beam, narrow in space and time.  
To obtain an excitation function the beam energy then had to be changed in
small steps of the order of the experimental resolution. 
The need for a radioactive target severely limits the 
applicability of this method to investigations in regions close to
$\beta$ stability.
However, it is possible to produce dripline species in 
simple reactions involving radioactive nuclei.
When using this approach,  
the beam is composed of radioactive ions and the target of light
nuclei, eliminating the need for a radioactive target.
Since this is the inverse setup to the one traditionally used in
scattering experiments, the method is usually denoted as elastic
scattering in inverse geometry. 

The advantage of using gas instead of a 
solid target is twofold. Firstly, the thickness of a gas target can be 
changed continuously and easily by adjusting the gas pressure, 
and secondly the target is very homogeneous. 
The beam parameters of radioactive ion beams (RIB's) are limited; 
the spread in both energy and space
are much larger than what can be obtained for stable beams, and
the intensities are of course much smaller.
As will be seen below, the beam properties are not of great 
importance in the experimental approach used here.
Elastic resonance scattering is
characterized by large cross sections and is therefore well suited
for use with low-intensity RIB's. 
These and several other features of the elastic
resonance scattering in inverse geometry on thick targets will be
illuminated  in the following subsections. 

The expressions for elastic cross section in the case of proton 
scattering on spin-less nuclei, eqs.~\ref{eq:dsdom} and~\ref{eq:parta}, 
can for example be found in~\cite{lau51:84b},
\begin{equation}\label{eq:dsdom}
\frac{d\sigma}{d\Omega}=\left|A(\theta)\right|^2+
\left|B(\theta)\right|^2,
\end{equation}
where
\begin{equation}\label{eq:parta}
\begin{split}
A(\theta)=\frac{zZ}{2\mu v\sin^2(\theta/2)}
e^{\frac{i\hbar}{\mu v}\ln \frac{1}{\sin^2(\theta/2)}}+ \\
\frac{1}{2ik} \sum_{\ell =0}^{\infty}
\left[(\ell+1)(e^{2i\delta^+_{\ell}}-1)+
\ell(e^{2i\delta^-_{\ell }}-1)\right]\\
e^{2i\sigma_\ell}P_\ell(\cos\theta)
\end{split}
\end{equation}
and
\begin{equation}\label{eq:par_b}
B(\theta)=
\frac{i}{2k}\sum_{\ell =0}^{\infty}
\left(e^{2i\delta^+_{\ell }}-e^{2i\delta^-_{\ell }}\right)
e^{2i\sigma_\ell}P^1_\ell(\cos\theta)\nonumber
\end{equation}
where $e^{2i\sigma_\ell}$ is defined by
\begin{equation}\label{eq:coul}
e^{2i\sigma_\ell}=\frac{\Gamma\left(\ell+1+\frac{i}{k}\right)}
{\Gamma\left(\ell+1-\frac{i}{k}\right)}
\end{equation}
Symbols introduced here are defined as:
\begin{tabular}{ll}
$^{\pm}$: & denotes states with $j$\ =\ $\ell\pm\frac{1}{2}$\\
$z$: & charge of the proton\\
$Z$: & charge of the spin-zero particle\\
$\mu$: & reduced mass\\
$v$: & the relative velocity of the particles\\
$k$: & the magnitude of the wave vector\\
$\sigma_\ell$: & the Coulomb phase shift\\
$P_{\ell}(cos\theta)$: & Legendre polynomial\\
$P^1_{\ell}(cos\theta)$: & associated Legendre polynomial
\end{tabular}

\vspace{3mm}
The first term in $A(\theta)$ represents the Coulomb 
scattering. The other terms in $A(\theta)$ and $B(\theta)$ express
scattering due to nuclear forces.
The phase shift $\delta_\ell$ is the sum of the phase shift from hard
sphere scattering, $-\phi_\ell$, and the resonant nuclear
phase shift, $\beta_\ell$,:
\begin{equation}\label{eq:delta}
\delta^{+}_{\ell}=\beta^{+}_{\ell}-\phi_{\ell},\quad
\delta^{-}_{\ell}=\beta^{-}_{\ell}-\phi_{\ell}
\end{equation}

The differential cross-section 
has its maximum in the vicinity of the position where 
the phase shift passes through $(n+\frac{1}{2})\pi$.
Therefore, a frequently used definition of the resonance energy
is where $\delta$\ = $(n+\frac{1}{2})\pi$, see section~\ref{sec:analysis}. 
It is favourable to study resonance scattering at 
180$^{\circ}$~c.m. where
eq.~\ref{eq:dsdom} is simplified. At this angle, only $m=0$
sub-states contribute to the cross section and both potential and
Coulomb scattering are minimal.
An advantage of the inverse geometry setup is its possibility to
measure at 180$^{\circ}$~c.m.

\subsection{Kinematical relations}
We define the laboratory 
energies of the bombarding particles before the interaction
in inverse (E) and conventional (T) geometry as $E_{0}$ and
$T_{0}$, respectively. The notation used mainly
follows ref.~\cite{mar68}, primed energies being in the c.m. system: \\ 
\\
\begin{tabular}{ll}
 $m$,$M$: & mass of the light and heavy particles \\
 $E_{M}$, $T_{M}$: & heavy particle $M$ laboratory energies \\
  &  after interaction   \\
 $E_{m}$, $T_{m}$: &
  light particle $m$ laboratory energies \\
  &  after interaction \\ 
 $\theta_{lab.}$: &  scattering angle of the light particle 
              in  \\
   & the laboratory system \\
\end{tabular}
\vspace{3mm}

\noindent 
The relations between laboratory energy of the beam and the 
c.m. energy of the heavy nucleus: \\ 
\begin{equation}
\label{eq:kn1}
{T_{M}^{'}}= T_{0}\frac{mM}{\left({M}+m\right)^2}  
\Leftrightarrow   
{E_{M}^{'}}={E_{0}\left(\frac{m}{{M}+m}\right)^2} 
\end{equation}
The expressions for the lab. energies of the light particle $m$ that
will be detected after scattering: \\
\begin{equation}
\begin{split}
\label{eq:kn2}
{T_{m}}={T_{0}}\Big(\frac{m}{{M}+m}\Big)^{2} 
\big(\cos{\theta_{lab.}}+\sqrt{K^{2}-\sin^{2}{\theta}}\big)^{2}\\
\Leftrightarrow
{E_{m}}={E_{0}}\frac{4mM}{({M}+m)^2}\cos^2\theta_{lab.} 
\end{split} 
\end{equation}
In the equation above, $K$ is the ratio of the masses
($E_{0}$/$T_{0}$=$M/m=K$ since ${E_{M}^{'}}={T_{M}^{'}}$).
Inserting $\theta_{lab.}= 0^{\circ}$ in eq.~\ref{eq:kn2} 
leads to the following ratio between 
the energy of the measured particle in conventional and inverse
geometry:
\begin{equation}\label{eq:kn4}
\frac{E_{m}}{T_{m}}=4\frac{K^2}{(1+K)^2} \sim 4
\end{equation}
\\
As is seen from eq.~\ref{eq:kn4}, the detected energy 
of the light particles is close to
4~times higher for inverse kinematics as compared to the
conventional geometry at the same c.m. energy. This is an
important gain for the study of resonant states near the threshold.
The excitation energy in the ${M}+m$ compound system
is obtained as the sum of the c.m. energies for particles $m$ and $M$:
\begin{equation}
\begin{split}
\label{eq:kn5}
{T_{ex}}={T_{0}}\frac{M}{{M}+m} 
\Leftrightarrow
{E_{ex}}={E_{0}}\frac{m}{{M}+m}
\end{split} 
\end{equation}
Using eq.~\ref{eq:kn2}, this can be expressed in terms of the measured
particle energy $E_m$. In case of inverse kinematics, the excitation
energy of the compound system becomes: 
\begin{equation}\label{eq:kin}
E_{ex} = \frac{{M}+m}{4M\cdot
  cos^{2}(\theta_{lab.})}E_{m}
\end{equation}
Because of the low energies involved, a non-relativistic expression can
be used.

\subsection{General set-up of elastic scattering in inverse geometry}
The basic experimental setup consists of a
radioactive ion beam which 
is incident on a scattering chamber filled with gas. 
The thickness of the gas target is adjusted to be 
slightly greater than the range of the beam ions.  
Charged-particle detectors are placed
at and around the beam direction,
i.e. 180$^{\circ}$~c.m., as shown in Figure~\ref{fig:setup}. 
As they are continuously slowed down in the gas, the
beam ions effectively scan the energy region from the
beam energy down to zero,
giving a continuous excitation function in this interval. 
When the energy of the heavy ion corresponds to a resonance in the
compound system, the cross section for elastic scattering increases 
dramatically and can exceed 1~b, making it possible to neglect the
non-resonant contributions which are on the order of mb. 
For the ideal case of a mono-energetic beam, each interaction point 
along the beam direction in the chamber corresponds uniquely to one 
resonance energy and, as we study elastic
scattering, to a specific proton energy for each given angle. 
Because the distance from the detector is different for each proton
energy, the solid angle also varies with proton energy and is
quite different for low and high energy resonances.

The high efficiency of the method 
is mainly a result of the large investigated
energy region. If we compare the
scanned region of 5-10~MeV with the typical energy step of 
$\approx$10-20 keV in conventional scattering measurements, the gain is
250-1000~times.  

\subsection{Energy resolution}\label{sec:e-resolution}
The initial energy spread of our $^{10}$C beam was 1.5\% 
of the total energy, which naturally increased along the beam path in
the gas. 
The energy spread of the beam results in excitation of the same
resonance at different distances from the detector. Assuming that
${\Delta}E$ is the energy spread at some point in the gas, this 
distance interval ${\Delta}x$ is given by
\begin{equation}
{\Delta}x = \frac{{\Delta}E}{(\frac{dE}{dx})_{M}}
\end{equation}
where $(\frac{dE}{dx})_{M}$ 
is the specific energy loss of the beam nuclei in the gas. 
Due to the protons energy loss in the gas, the measured proton energies 
corresponding to the same resonance are slightly different. 
The resulting spread of proton energies, ${\varepsilon}$, corresponding
to the interval ${\Delta}x$ will be 
\begin{equation}
\varepsilon={\Delta}E \frac
{(\frac{dE}{dx})_{m}}
{(\frac{dE}{dx})_{M}}
\end{equation}
Here, $(\frac{dE}{dx})_{m}$ denotes the specific energy loss of the
recoil nuclei (protons) in the gas.
Taking into account the different velocities of the beam ions and the
scattered protons as well as the Bethe-Bloch 
expression for specific energy loss, one finds:
\begin{equation}\label{eq:elos} 
{\varepsilon}{\sim}\frac{{\Delta}E}{4}\frac{z^2}{Z^2}
\end{equation}
In the case of $^{10}$C+p interaction, eq.~\ref{eq:elos} becomes 
${\varepsilon}{\sim}\frac{{\Delta}E}{144}$.
Hence, for $\Delta$E~=~5~MeV a lab. energy resolution of 35 keV is
expected. The effective c.m. energy resolution will be about four  
times better than the resolution in the lab. frame, see eq.~\ref{eq:kn4}.
Thus it is clearly shown that the energy spread of the radioactive beam
does not restrict the applicability of the method.
Many other factors influence the final resolution, for example the
size of the beam spot and detectors, the detector
resolution, the angular divergence of the beam and straggling of 
light particles in the gas. These factors can be taken into account
by Monte Carlo simulations. 
In reality, an effective energy resolution of 20~keV in the c.m. frame 
is feasible. At angles other than 180$^{\circ}$ the resolution
deteriorates,
mainly due to kinematical broadening of the energy signals for protons 
scattered at different angles. This contribution to the 
resolution could be reduced by tracking the proton angles.

\subsection{Background sources}\label{sec:thback}
A cornerstone of the described experimental approach is that 
elastic resonance scattering dominates over other possible processes.
The competing reaction channel which has to be treated for
each specific case is inelastic resonance scattering, 
as it is a 
resonant process which produces the same recoil particles as the 
elastic scattering. However, the elastic and inelastic resonance 
scattering reactions can be distinguished from each other.
The energy of the scattered nuclei from inelastic resonance
scattering at~0$^{\circ}$ is given by eq.~\ref{eq:e_light} if 
$E^{*}/E_{0} {\ll}$ 1, where 
$E^*$ is the excitation energy of the beam nucleus~\cite{mar68}.  
\begin{equation}\label{eq:e_light}
E_{m}\ {\approx}\ 4\frac{mM}{({M}+m)^2}
\Big(E_{0}-{\frac{E^*}{2}}{\frac{{M}+m}{m}}\Big)
\end{equation}
Comparing this with eq.~\ref{eq:kin}, one sees that
the energy of heavy ions has to be 
larger by an amount $\mathcal{E}$ for the inelastic scattering
to obtain the same energy of a light recoil from 
the elastic and inelastic scattering reactions, when $\mathcal{E}$ is
defined in eq.~\ref{eq:kinema} . 
\begin{equation}\label{eq:kinema}
 \mathcal{E} {\approx} \frac{\mathnormal{E}^*}{\mathrm{2}}
 \mathnormal{{\frac{{M}+m}{m}}}
\end{equation} 
For the $^{10}$C+p case, where $E^*$($^{10}$C(2$^+_1$)) = 3.35~MeV,
eq.~\ref{eq:kinema} shows that the inelastic resonance 
scattering should take place at about 20~MeV higher energy 
than the elastic one for the two processes to mix in the elastic
scattering excitation function.
The inelastic resonance reaction thus has to take place 
further from the detectors, closer to the entrance window, 
in order to produce a scattered particle with the same
energy as the corresponding elastic process.
The two processes in question hence can give the same energies of
the recoil protons but their ToF (window-detector) will differ.
The time difference between the two types of events will be on the order
of a few ns, and can thus be separated in the analysis. No
such events were seen in our data.

Other scattering reactions contribute very little to the spectrum,
especially at 180$^{\circ}$~c.m., the exception being 
low energies where the Coulomb
scattering cross sections increases. However, this scattering is well
understood and can be included in the data treatment.
Additional sources of background are $\beta$ particles from decaying
radioactive ions in the gas, beam ions which penetrate the gas target, 
and particles scattered in the entrance window. 

\section{Experimental procedure}\label{sec:expproc}
\begin{figure}[htb]
  \centerline{\epsfig{file=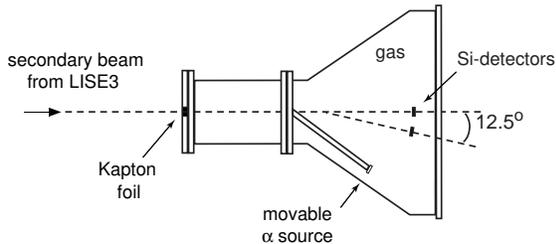, width=0.95\linewidth}}
  \caption{A schematic view of the scattering chamber and
    Si-detectors. The 12.5$^{\circ}$ detector has this angle to the
    middle of the chamber. The
    interaction position and the corresponding angle is calculated for
    each event when analyzing the data. }
  \label{fig:setup}
\end{figure}

The first experiment was performed using the LISE3 spectrometer 
at the GANIL heavy-ion facility. The secondary $^{10}$C beam 
was produced by a 75~MeV/u $^{12}$C$^{6+}$ beam with an 
intensity of 2$\cdot$10$^{12}$~ions/s which bombarded a 
8~mm thick, rotating Be target and a fixed 400~$\mu$m Ta target. 
The $^{10}$C fragments were selected 
in the LISE3 spectrometer, using an achromatic 
degrader at the intermediate focal plane (Be, 220 $\mu$m thick) and the 
Wien-filter after the last dipole. 
The 50~cm long scattering chamber was placed at the final 
focal plane. Immediately before the 80~$\mu$m thick kapton 
entrance window, a PPAC (Parallel
Plate Avalanche Counter) registered the position of the incoming ions.
The intensity of the secondary beam, measured by the 
PPAC, was approximately 7000~ions/s, and due to the degrader and 
Wien filter a very low degree of contamination was achieved.  
The efficiency of the PPAC at this intensity and
ion charge is close to 100\%, which makes it easy to use the
PPAC count rate to obtain absolute cross-sections.
 The scattering chamber was filled with CH$_{4}$ gas, acting as a thick
proton target for the incoming $^{10}$C ions. 
The gas pressure was adjusted to 816$\pm$5~mbar, 
which was the pressure required
to stop the incoming beam just in front of the central 
detector.
It is desirable to stop the beam close to the detectors in order to 
avoid loosing any protons scattered from a possible 
low-lying resonance in $^{11}$N.
In the far end of the chamber
an array of Si-detectors was placed. 
The detectors had diameters of 20~mm and thickness of 2.50~mm,
corresponding to the range of 20~MeV protons.
The time  between the radio frequency (RF)
from the cyclotron and the PPAC gave one time-of-flight signal 
(ToF), while the time difference of the PPAC and detector signal 
gave additional ToF-information. 
The complete setup is shown in Figure~\ref{fig:setup}.

As a first measurement, a low intensity $^{10}$C beam 
was sent into the evacuated scattering chamber
to get the total energy and spread of the secondary beam after the
foil, and this was determined to be 90~MeV with a FWHM~=~1.5~MeV.
For background measurements, the scattering chamber was filled with
CO$_{2}$ gas at 450$\pm$5~mbar and bombarded with $^{12}$C and $^{10}$C
beams, respectively. 
For our purposes, we assume that $^{16}$O and $^{12}$C behave
similarly in proton scattering reactions.
The measurements with the CO$_{2}$ target would reveal any background
stemming from the carbon nuclei in the CH$_4$ target 
gas or from the kapton window.
Beam contaminations would also be present in
these runs, and those background sources can subsequently 
be subtracted from the experimental excitation functions. 
The standard beam diagnostics observed admixtures of d, $\alpha$
and $^{6}$Li with the same velocity as the $^{12}$C secondary 
beam, while no contaminant particles could be seen in the $^{10}$C beam.
The $^{10}$C+CO$_2$ spectrum showed no prominent structure and
was found to contribute less than 10\% to the total cross section.
This background spectrum was subtracted from the $^{10}$C+CH$_4$
spectrum before transformation to the c.m. system.

\begin{figure}[htb]
 \centerline{\epsfig{file=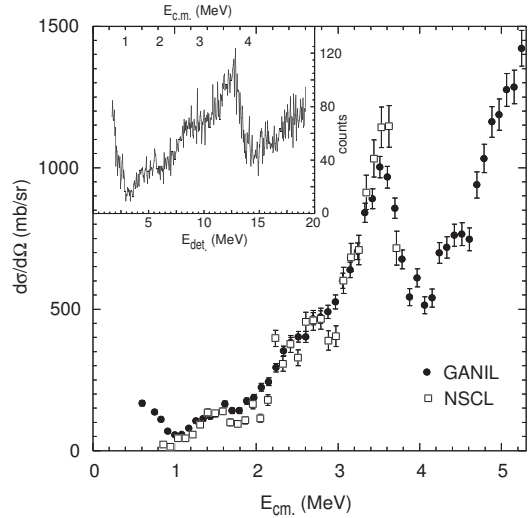, width=0.9\linewidth}}
 \caption{The excitation function of $^{11}$N is shown 
  after background subtraction
  and transformation to the c.m. system. The filled black circles
  represent the GANIL data, and the white squares show the result of the
  MSU experiment. The inset shows the raw data in the lab. system. 
  The upper scale in the inset is a rough calibration
  to c.m. energy, shown for comparison, while the lower scale is the
  detected proton energy.}
  \label{fig:nofit}
\end{figure}

Since $^{10}$C is a $\beta^{+}$ emitter
with a half life of 19.3~s, it is necessary to discriminate
the positron signals from the protons.
This was done by selecting the protons in a two-dimensional
spectrum showing ToF (PPAC to Si-detector) versus detected energy,
where the positrons are clearly distinguished from protons both by 
their uniform time distribution and their maximum energy of
1.93~MeV. A positron with energy in this interval
has a maximum energy loss of 1.25~MeV in 2.50~mm Si, 
which simulates a scattered proton energy of 0.344~MeV in the 
excitation function of $^{11}$N.
Since the positron energies are small enough to lie 
in the energy range of Coulomb scattered protons, cutting away all
events below this energy does not distort the interesting parts of the
proton spectrum, as is seen in the inset in Figure~\ref{fig:nofit}.

In this paragraph we justify our ignoring the
background contributions to our spectra from inelastic 
scattering of $^{10}$C on hydrogen with excitation 
of the particle stable 2$^+$ level at 3.35~MeV in $^{10}$C.
The contribution from inelastic scattering 
has been estimated using available data on inelastic scattering of
protons on a $^{10}$Be 
target~\cite{aut70:157} and a DWBA extrapolation to the whole investigated
interval of energies. This shows that the contribution from inelastic
scattering does not exceed 1\% of the observed cross section.

\begin{figure}[htb]
  \centerline{\epsfig{file=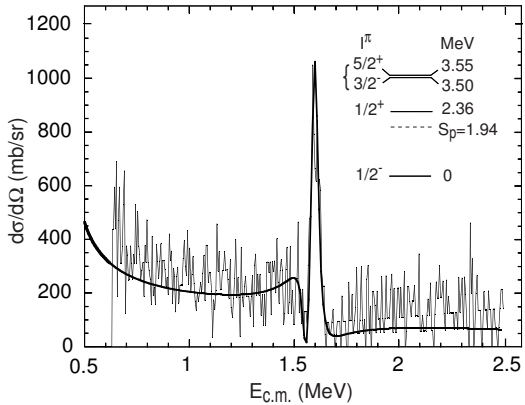, width=0.9\linewidth}}  
  \caption{Experimental spectrum of protons from the scattering of
    $^{12}$C. The energy is given as c.m. energy of
    $^{13}$N. The peaks are fitted with two coherently added
    resonances as described in the text. }
  \label{fig:13n}
\end{figure}

The energy calibration of the Si detectors 
was done with a triple $\alpha$-source 
($^{244}$Cm, $^{241}$Am and $^{239}$Pu) which was 
placed on a movable arm inside the chamber.
Another calibration, and at the same time a performance check 
of the setup, was 
done by investigating known resonances in $^{13}$N.
The primary $^{12}$C beam, degraded to 6.25~MeV/u, was
scattered on the methane target using a gas pressure of 240$\pm$5~mbar. 
The resulting proton spectrum clearly shows
the two closely lying resonances in $^{13}$N 
(3.50~MeV, width 62~keV, and 3.55~MeV,
width 47~keV,~\cite{jac53:89}), as can be seen in Figure~\ref{fig:13n}. 
These resonances are overlapping and the width of the peak is 50~keV.  
The solid curve in Figure~\ref{fig:13n} is a fit obtained by coherently
adding two curves in order to take interference into account. 
The 5/2$^{+}$ resonance at 3.55~MeV has single particle (SP)
nature~\cite{jac53:89}  
and was described using the potential model outlined in 
section~\ref{sec:analysis}, while a Breit-Wigner curve was used for 
the 3/2$^{-}$ state at 3.50~MeV.
The resonant 1/2$^+$ state in $^{13}$N, 420~keV above
the $^{12}$C+p threshold, is not seen as it is overlapping the
Coulomb scattering which dominates 
below 0.5~MeV. 
From the calibration measurements described above an energy 
resolution of 100~keV in the lab. frame was deduced, mainly
determined by the detector 
resolutions and proton straggling in the gas.

The experimental proton spectrum was, after subtraction of the measured
carbon background, transformed into 
differential cross-section as a function of the excitation 
energy of $^{11}$N, 
in the following referred to as the excitation function of $^{11}$N.
Since each interaction point along the beam direction ideally 
corresponds to a
specific resonance energy, the measured proton energy can, after
correction for its energy loss in the gas, be used to find the
resonance energy in $^{11}$N in the c.m. system.
The inset in Figure~\ref{fig:nofit} shows the experimental data as
measured proton energy versus counts at~0$^{\circ}$ before the corrections 
for solid angle was made. 
Comparing this pictures to the one obtained after transformation to
the c.m. system clearly shows 
the effect of differing solid angles for
different proton energies. The cross sections in the
high energy part increases relative to the low energy part, as is
clearly seen when comparing the inset to the transformed spectrum in
Figure~\ref{fig:nofit}.  

Extracting the cross section from the data is
straightforward, and the transformation to c.m. is done using
eq.~\ref{eq:cross3}. 
\begin{equation}
\label{eq:cross3}
 \Big(\frac{d\sigma}{d\Omega}\Big)_{c.m.} =
\frac{1}{4cos(\theta_{lab.})}\Big(\frac{d\sigma}{d\Omega}\Big)_{lab.} 
\end{equation}
The relation between the
scattering angle in the lab. and c.m. systems is simply
$\theta_{c.m.}$=180$^\circ$-2$\theta_{lab.}$. 
The excitation function obtained after background subtraction and
transformation into the c.m. system is shown in Figure~\ref{fig:nofit}.

The more detailed analysis now performed revises the absolute cross
section to a larger value from what was previously published in~\cite{axe96:54}.

A second independent measurement of $^{10}$C+p
elastic scattering using the same method was made at NSCL where the A1200
spectrometer delivered the $^{10}$C beam.
The experimental conditions were the same as in the GANIL experiment, 
except that at NSCL a $\Delta$E-E telescope was placed at 0$^{\circ}$ 
and no Wien-filter was used. 
The energy of the $^{10}$C beam after the foil was 7.4~MeV/u and the
beam intensity was 2000~pps.
The data from these two experiments are
overlaid in Figure~\ref{fig:nofit} where it is seen that the 
structures and the absolute cross sections coincide. 

\section{Analysis and results}\label{sec:analysis}
The excitation function, shown in Figure~\ref{fig:nofit}, reveals
structure in the region from 1~MeV to 4~MeV that could be due to
interfering broad resonances. A reasonable first assumption is that the
structure corresponds to the three lowest states in $^{11}$N.
This assumption is justified by the closed proton $p_{3/2}$ sub-shell in
$^{10}$C and agrees with the known predominantly 
single particle nature of the lowest states in $^{11}$Be~\cite{ajz90:506}, 
the mirror nucleus of $^{11}$N.
Taking this as a starting point, we assume that the observed levels in
$^{11}$N are mainly of SP nature.
The SP assumption validates the use of a shell-model potential to describe the
experimental data of $^{11}$N.

\subsection{Analysis of the three lowest levels in $^{11}$N}\label{sec:potmod}   
The $^{11}$N states are all in the continuum and the aim of the analysis
was to obtain $I^{\pi}$ and other resonance parameters as it can 
be done in the framework of the optical model. 
Because of the absence of other scattering processes than the 
elastic scattering channel, no imaginary part is included in the
potential.
The potential has a common form consisting of a Woods-Saxon central
potential and a spin-orbit term with 
the form of a derivative of a Woods-Saxon potential. 
The Woods-Saxon ($\ell$s) term has the usual parameters
$V_0$ ($V_{{\ell}s}$), $r_0$ ($r_{{\ell}s}$) and $a_0$ ($a_{{\ell}s}$)
for well depth, radius and diffusity, respectively.
Centrifugal and Coulomb terms were also included in the potential. 
The Coulomb term has the shape of a uniformly charged sphere with radius
$r_{c}$.  
The full potential is given in eq.~\ref{eq:pot}, where $\mu$ is the
reduced mass of the system and $\lambda_{\pi}$ denotes the  pion Compton
wavelength. 

\begin{equation}\label{eq:pot}
\begin{split}
V_{0} = \frac{V_{\ell}}
{1+e^{\frac{r-R_0}{a_0}}}+
{\bf \ell\cdot s}\frac{V_{\ell s}(\lambda_{\pi}/2\pi)^2}{a_{\ell s}}\times\\
\frac{e^{\frac{r-R_{\ell s}}{a_{\ell s}}}}
{\left(1+e^{\frac{r-R_{\ell s}}{a_{\ell s}}}\right)^2}+
\quad \frac{\ell(\ell+1) \hbar^{2}}{2{\mu} r^{2}}+V_{c}
\end{split}
\end{equation}
\begin{eqnarray}
V_{c}& =& \begin{cases} 
 \frac{zZ}{2} 
 \frac{e^{2}}{4{\pi}{\epsilon_{0}} R_{c}}  
 \left(3 - \frac{r^{2}}{R_{c}^{2}} \right)   
  &:\quad r < R_c \nonumber   \\
  \frac{zZe^{2}}{r}   
  &:\quad r > R_c \nonumber     
\end{cases}
\end{eqnarray}
\begin{equation}
R_0=r_0A^{1/3},\ 
R_{c}=r_{c}A^{1/3},\ 
R_{\ell s}=r_0A^{1/3}\nonumber
\end{equation}

As a starting point, standard values of the potential parameters were
chosen~\cite{kra88} and the well depths were varied separately for
each partial wave ($\ell$=0,1,2), see Table~\ref{tab:para}:a.
The cross-section of the experimental data at 180$^\circ$
was found to be larger than predicted by the potential model. 
As can be seen in the experimental spectrum, Figure~\ref{fig:nofit},
there is substantial amount of cross section above 4~MeV, indicating
additional resonances in this energy region.
\begin{figure}[htb]
 \centerline{\epsfig{file=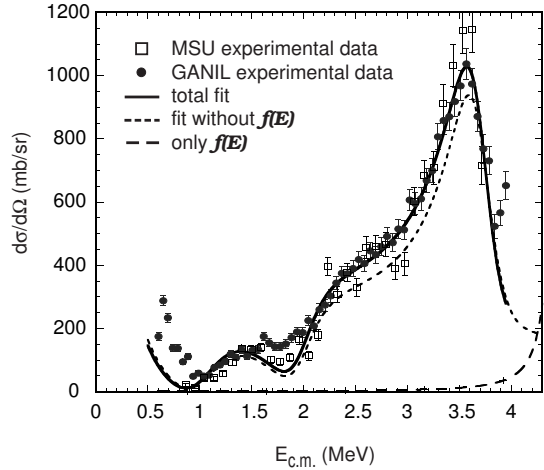, width=0.93\linewidth}}
  \caption{Experimental excitation function of $^{11}$N together with
    the best fit from the SP-model program. The energy is given as 
           excitation energy of $^{11}$N. The fit is made using a
           potential with parameters shown in Table~\ref{tab:para}:c.
           The underestimation of the cross section around 1.8~MeV is
           the only failure of the potential model. This part is
           better described when the influence of higher resonances
           are taken into account, see Figure~\ref{fig:rmatrix}.}
  \label{fig:expspe}
\end{figure}
The underestimation of the potential model can thus be attributed to 
influence of higher lying resonances.
To simulate the presence of those highly excited states, an
amplitude $f$ was added to the amplitude calculated from 
the potential model. The form of this extra
amplitude was $f=\frac{b}{E_0-E}$, where $E_0$ was taken as a constant
(4.5~MeV) and $b$ was used as a parameter.
As is seen in Figure~\ref{fig:expspe}, the introduced amplitude is
small in comparison with the measured cross sections, but it nonetheless
was useful in the fitting procedure. 
A more sophisticated way to include
the influence of higher resonances is to use an R-matrix procedure, and
some attempts in this way were also made, see sec.~\ref{sec:rmatrix}.

The best fit
for conventional parameter values, only varying $V_\ell$ is 
obtained for the level ordering $1s_{1/2}$,
$0p_{1/2}$, and $0d_{5/2}$, and the corresponding parameters are given in
Table~\ref{tab:para}:a. 
The curve resulting from these parameters does not differ
  significantly from the one obtained using the parameter set
  Table~\ref{tab:para}:c.

\begin{figure}[htb]
  \centerline{\epsfig{file=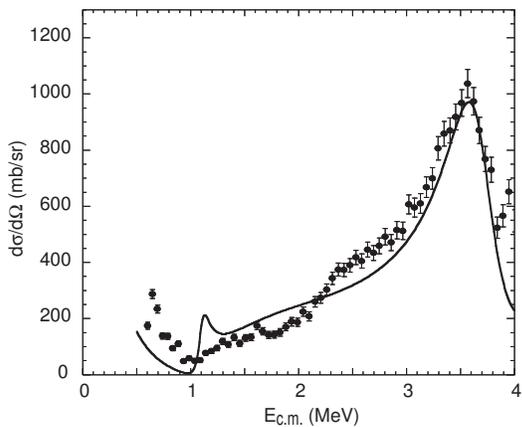, width=0.9\linewidth}}  
  \caption{A typical fit of the experimental data when the
    1/2$^{-}$ state is assumed to be the ground state of $^{11}$N. It
    is evident that the excitation function is not well
    described. Putting the $p$ state this low makes it too narrow, at
    the same time as the $s$ state becomes very broad since it is now
    well above the Coulomb barrier.}
  \label{fig:nopi}
\end{figure}

\begin{figure}[htb]
  \centerline{\epsfig{file=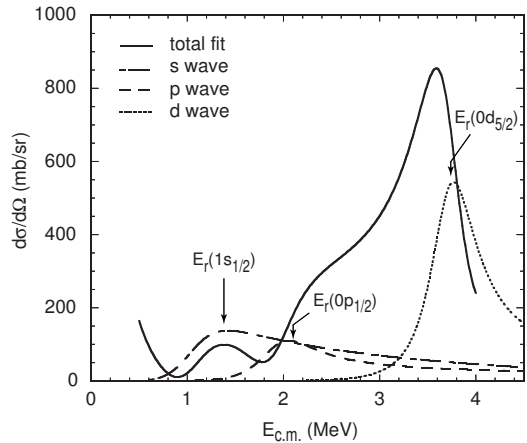, width=0.9\linewidth}}
  \caption{The partial waves $s_{1/2}$, $p_{1/2}$ and $d_{5/2}$
    together with the total calculated excitation function for the best
  fit to the experimental data (Table~\ref{tab:para}:c).}
  \label{fig:pw}
\end{figure}

\begin{figure}[htb]
   \centerline{\epsfig{file=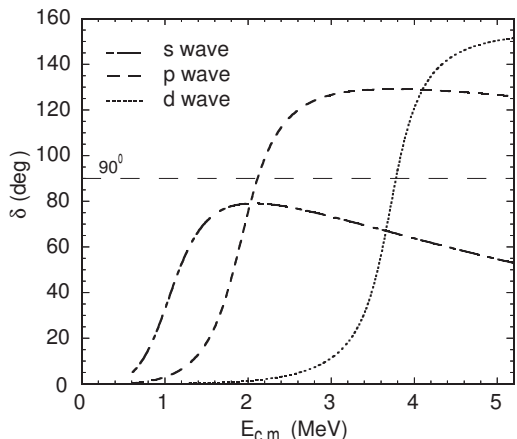, width=0.9\linewidth}}  
  \caption{The phase shifts from the theoretically calculated curve
    which is the best fit to the experimental data 
        (Table~\ref{tab:para}:c).}
  \label{fig:ps}
\end{figure}

A potential with conventional parameters and the same well depth for all
$\ell$ will generate
single particle levels in the order $0p_{1/2}$, $1s_{1/2}$ and
$0d_{5/2}$ above the $0p_{3/2}$ sub-shell.   
However, all attempts to describe the experimental data keeping this
ordering of the levels failed. A typical example of a calculated
excitation function with this level sequence is shown in
Figure~\ref{fig:nopi}, with parameters in Table~\ref{tab:para}:b.
This result is not surprising when considering the well known level
inversion in $^{11}$Be.
For the potential in Table~\ref{tab:para}:c, the cross section
of each partial wave is shown in Figure~\ref{fig:pw} together 
with the total calculated curve. Comparing the partial cross sections
with the total cross section, it is clear that interference between
the partial waves determine the shape of the total curve.
The corresponding phase shifts are shown in Figure~\ref{fig:ps}.
The most common definition of resonance energy is where the phase
shift $\delta_{\ell}$ passes $\pi$/2. As is seen in Figure~\ref{fig:ps}, the
phase of the 1/2$^{+}$ resonance, which is the broadest level, does not reach
$\pi$/2. 
Therefore, we have defined the resonance
energy as where the partial-wave amplitude 
calculated at r~=~1~fm has its maximum.
The width is defined as the FWHM of the partial wave.
One can note that for the 1/2$^{-}$ and 5/2$^{+}$ levels, the same
resonance energies are 
obtained by our definition and $\delta_{\ell}$~=~$\pi$/2.
All attempts to change the resonance spins and parities or the order of
the levels resulted in obvious disagreement with the experimental
data. 
We thus conclude that the unambiguous spin-parity assignments for the lowest
states in $^{11}$N are a 1/2$^{+}$ ground state, a first excited
1/2$^{-}$ level and a 5/2$^{+}$ second excited state. 
All further fitting procedures were performed
with the aim to obtain more exact data on the positions and widths of 
the resonances.

A disadvantage of the potential model is that it produces 
resonances with single-particle widths. In general, 
the nature of the states is more complicated and their widths can be
smaller than what is predicted by the potential model. 
To investigate how changing the resonance widths would affect the
overall fit, we 
changed the radius parameter $r_0$, and fitted new well depths to
get the best possible agreement with the data.
It was evident that the widths obtained with $r_0$~=~1.4~fm are too large
for the 1/2$^{-}$ and 5/2$^{+}$ resonances, while $r_0$~=~1.0~fm makes
these levels too narrow. 
\begin{table}
\caption{The sets of potential parameters used to obtain the best fits 
          of the experimental data, and the resulting resonance parameters.
$V_{\ell s}$=5.5~MeV is kept the same in all fits. The change
  of this parameter gave only minor modifications. The parameter $b$
  given in the footnotes is the value used in $f(E)$=b/(4.5-E). }
\begin{center}
\begin{tabular}{cc|cccc|cc}
& & \multicolumn{4}{c|}{Potential parameters}  &  \multicolumn{2}{c}{Resonance} \\
& & $V_\ell$ &$r_0$, $r_{\ell s}$, $r_{c}$ & $a_0$ &
$a_{\ell s}$ & $E_r$ & $\Gamma_r$ \\
&& (MeV)    & (fm)  &       (fm)    & (fm)  & (MeV) &  (MeV)\\
\hline
$^{a}$
      & $1s_{1/2}$ &-66.066  &1.20  &0.53  &0.53 &1.30  &1.24  \\
      & $0p_{1/2}$ &-42.336  &1.20  &0.53  &0.53 &1.96  &0.65   \\
      & $0p_{3/2}$ &-42.084  &1.20  &0.53  &0.53 &-1.06  & -    \\
      & $0d_{3/2}$ &-78.792  &1.20  &0.53  &0.53 &4.40  &0.90   \\
      & $0d_{5/2}$ &-64.092  &1.20  &0.53  &0.53 &3.72  &0.61    \\\hline
$^{b}$
      & $1s_{1/2}$ &-45.360  &1.40  &0.65  &0.30  &1.70  &3.49    \\
      & $0p_{1/2}$ &-33.474  &1.40  &0.28  &0.30  &1.11  &0.11   \\
      & $0p_{3/2}$ &-32.340  &1.40  &0.53  &0.30  &-1.22  &-   \\
      & $0d_{3/2}$ &-58.086  &1.40  &0.53  &0.30  &4.45  &1.23    \\
      & $0d_{5/2}$ &-45.570  &1.40  &0.35  &0.30  &3.75  &0.60    \\\hline
$^{c}$
      & $1s_{1/2}$ &-47.544  &1.40  &0.65  &0.30  &1.27  &1.44    \\
      & $0p_{1/2}$ &-31.500  &1.40  &0.55  &0.30  &2.01  &0.84    \\
      & $0p_{3/2}$ &-32.592  &1.40  &0.53  &0.30  &-1.33  &-    \\
      & $0d_{3/2}$ &-57.960  &1.40  &0.53  &0.30  &4.5  &1.27    \\
      & $0d_{5/2}$ &-45.570  &1.40  &0.35  &0.30  &3.75  &0.60     \\\hline
$^{d}$
      & $1s_{1/2}$ &-56.280  &1.20  &0.75  &0.30  &1.32  &1.76    \\
      & $0p_{1/2}$ &-42.420  &1.20  &0.55  &0.30  &2.14  &0.88    \\
      & $0p_{3/2}$ &-42.210  &1.20  &0.53  &0.30  &-1.33  &-    \\
      & $0d_{3/2}$ &-78.960  &1.20  &0.53  &0.30  &5.0  &1.39    \\
      & $0d_{5/2}$ &-62.874  &1.20  &0.50  &0.30  &3.79  &0.59    \\\hline
$^{e}$
      & $1s_{1/2}$ & -66.066 & 1.20 & 0.53 & 0.53 & 1.30  & 1.24 \\
      & $0p_{1/2}$ & -42.084 & 1.20 & 0.53 & 0.53 & 2.04  & 0.72 \\
      & $0p_{3/2}$ & -42.084 & 1.20 & 0.53 & 0.53 & -1.06 & -  \\
      & $0d_{3/2}$ & -64.092 & 1.20 & 0.53 & 0.53 & 9.87  & 4.53  \\
      & $0d_{5/2}$ & -64.092 & 1.20 & 0.53 & 0.53 & 3.72  & 0.61 \\\hline
$^{f}$
      & $1s_{1/2}$ &-46.074  & 1.40 & 0.70 & 0.30 &1.27  &1.56  \\
      & $0p_{1/2}$ &-30.492  & 1.40 & 0.70 & 0.30 &2.01  &1.09  \\
      & $0p_{3/2}$ &-32.550  & 1.40 & 0.53 & 0.30 &-1.31 & -  \\
      & $0d_{3/2}$ &-57.960  & 1.40 & 0.53 & 0.30 &4.50  &1.27   \\
      & $0d_{5/2}$ &-42.378  & 1.40 & 0.70 & 0.30 &3.75  &1.08  \\
\hline\hline
\multicolumn{8}{l}{\footnotesize $^{a}$ $r_0$=1.2~fm, only varying
      $V_\ell$ 
(\textit{b}=1.25)}\\
\multicolumn{8}{l}{\footnotesize $^{b}$ No level inversion (\textit{b}=1.25)}\\
\multicolumn{8}{l}{\footnotesize $^{c}$ $r_0$=1.4~fm, varying $a$ and
      $V_\ell$ (\textit{b}=1.25):}\\
\multicolumn{8}{l}{\footnotesize \hspace{1mm} the best fit to the data} \\
\multicolumn{8}{l}{\footnotesize $^{d}$ $r_0$=1.2~fm, varying $a$ and
      $V_\ell$ 
  (\textit{b}=2.4)}\\
\multicolumn{8}{l}{\footnotesize $^{e}$ The parameters used
  in~\cite{axe96:54} (\textit{b}=0)}\\
\multicolumn{8}{l}{\footnotesize $^{f}$ The parameters used to obtain the widths in} \\
\multicolumn{8}{l}{\footnotesize \hspace{1mm} the single particle limit.} \\
\end{tabular}
\end{center}
\label{tab:para}
\end{table}
As the 1/2$^{+}$ state is least affected by the
change of radius the conclusion for this level is difficult, but the
largest obtained width seemed most appropriate.
Therefore, the radii parameters $r_0$, $r_{{\ell}s}$ and $r_c$ were
chosen as 1.4~fm, 
and the well depths and diffusities were varied separately for each
$\ell$ to obtain the best fit of the experimental data up to 4~MeV.
An additional argument for choosing the larger radius was the fact that
this parameter value gives a good simultaneous description of the mirror
pair $^{11}$Be and $^{11}$N~\cite{axe96:54}.
The curve obtained in this way
that agreed best with the experimental excitation function 
is shown in Figure~\ref{fig:expspe}, and the 
corresponding potential parameters and resonance energies 
and widths are shown in Table~\ref{tab:para}:c.
For comparison, the values for $r_0$=$r_{{\ell}s}$=$r_c$=1.2~fm are
also given in Table~\ref{tab:para}:d. 

The extracted resonance parameters show a remarkable stability against
changes in the potential parameter sets, meaning that different sets of
parameters that fit the data give similar resonance energies and widths. 
This is seen in Table~\ref{tab:para}, comparing different sets of
parameters. The final energies
and widths are listed in table~\ref{tab:comp}. The error bars include
systematic errors and are dominated by a contribution from the spread 
in results from different parameter sets. Contributions from background
subtraction and solid angle corrections will be much smaller than those
sources. 

The SP reduced widths could be extracted 
for the three lowest levels where the only possible decay channel
is one-proton emission to the ground state of $^{10}$C.
The values of reduced widths are usually presented as a
ratio to the Wigner limit, which serves as a measure of the single
particle width~\cite{bre36:49}.
In our case we have a way to give a more exact 
evaluation of the reduced widths as the ratio of the 
widths obtained from the shell-model potential that fits the data
(Table~\ref{tab:para}:c)
to the widths calculated from a true shell-model potential. 
These ratios are free from the uncertainties related with
different definitions of the level widths.

Since the true shell model potential is not known for
$^{11}$N, and we approximated this potential with the parameters
shown in Table~\ref{tab:para}:f.
\begin{table}[htb]
 \caption{Reduced widths for the observed states obtained from the
   ratio of the widths from
   parameter sets c (experimental widths) and f (single particle
   widths) in Table~\ref{tab:para}.}
 \begin{tabular}{c|c}
\hspace{7mm} Level I$^{\pi}$ \hspace{1cm}& Reduced width $\Gamma_{exp}$/$\Gamma_{SP}$\\
\hline 
\hspace{1cm} 1/2$^+$ \hspace{1cm} & 0.92$\pm$0.2 \\
\hspace{1cm} 1/2$^-$ \hspace{1cm} & 0.77$\pm$0.2\\
\hspace{1cm} 5/2$^+$ \hspace{1cm} & 0.56$\pm$0.2\\
\hspace{1cm} 3/2$^{-}$ \hspace{1cm} & 0.15$\pm$0.2 \\
 \end{tabular}
\label{tab:sf}
\end{table}

Justification for using this particular set as shell model
potential is that it 
simultaneously reproduces the level positions in both $^{11}$Be and
$^{11}$N and gives a width of the 1/2$^+$ state that is larger than
for the parameters in Table~\ref{tab:para}:c. 
The reduced widths obtained in this way are given in Table~\ref{tab:sf}.

\begin{figure}[htb]
 \centerline{\epsfig{file=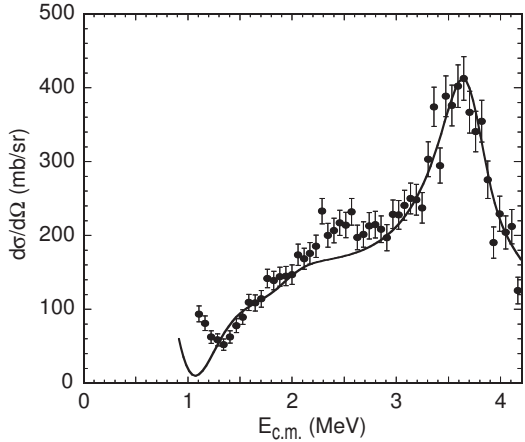, width=0.9\linewidth}}
  \caption{Experimental spectrum of protons in the detector placed at 
 12.5$^{\circ}$ relative to the center of the chamber. The full drawn
    curve is the result from the potential model using the parameters
    given in Table~\ref{tab:para}:c.}
  \label{fig:exp125}
\end{figure}

The potential parameters for the fit of the data at 180$^{\circ}$
(parameter set~c in~\ref{tab:para})  
were used to describe the excitation function obtained by a
detector at 12.5$^{\circ}$ relative to the center of the chamber.
The experimental data from this case are shown in Figure~\ref{fig:exp125}
together with the theoretical curve without the scaling amplitude $f$.
Comparing the experimental excitation functions in Figures~\ref{fig:expspe}
and~\ref{fig:exp125}, rather big differences are seen. 

\noindent
This reflects the fact that the laboratory angle depends on the interaction
point in the chamber. The angular range goes from
$\theta_{c.m.}$=150$^{\circ}$ for protons from higher resonances to
$\theta_{c.m.}$=93$^{\circ}$ for low energy protons.
This is taken into account in the calculation of the
excitation function, and from Figure~\ref{fig:exp125} it is evident that
the potential model describe the observed changes in the excitation
function with angle, a fact which supports our interpretation. 

\subsection{Analysis of the full excitation function}\label{sec:rmatrix}
In an attempt to investigate the influence of higher lying states on the
cross section in the lower part of the experimental spectrum, 
a simplified R-matrix approach was used.
For $^{11}$Be, about 10 levels are predicted in the energy region
2.7~MeV to 5.5~MeV~\cite{liu90:42}, but only four resonances have been
experimentally found~\cite{ajz90:506}.
The knowledge of the levels in $^{11}$N is even more incomplete, and our 
experimental data are not sufficient for a detailed R-matrix analysis.
Therefore, the treatment described below was performed rather 
to outline possible questions than to give definite answers.
The analysis was made using the potential model and adding
resonances at energies above 4~MeV according to eq.~\ref{eq:rmx}.

\begin{equation}\label{eq:rmx}
\begin{split}
\frac{d\sigma}{d\Omega} \bigg(\theta=180^{\circ}\bigg)= \\
\Bigg|A_{pot}-\frac{i}{2k} \sum_{n_{\ell}}^{+ } \bigg[({\ell}+1)\ 
\left(e^{2i\beta_{\ell}^{+}}-1\right)
\ e^{2i(\phi_{\ell}^{+}+\sigma_{{\ell}})}\bigg] \\
-\frac{i}{2k} \sum_{n_{\ell}^{-}}^{ } \bigg[{\ell}\ 
\left(e^{2i\beta_{\ell}^{-}}-1\right)
\ e^{2i(\phi_{\ell}^{-}+\sigma_{{\ell}})}\bigg]\Bigg|^2
\end{split}
\end{equation}

Two known levels in $^{11}$Be (2.69~MeV, $\Gamma$~=~200~keV, 
and 3.41~MeV, $\Gamma$~=~125~keV) were taken into account. 
The energy of those resonances in $^{11}$N were determined by
calculating the Coulomb differences 
between the mirror nuclei using the potential model. 
To fit the experimental data, the resonance energies were varied around
the value determined from the Coulomb-energy calculation. The values
finally used in the R-matrix fit are shown in Table~\ref{tab:rmatrix}. 
\vspace{3mm}

\begin{tabular}{ll}
$^\pm$: & stands for states with $j=\ell \pm$1/2, resp.\\
$A_{pot}$: & potential model amplitude at 180$^{\circ}$,\\
 &  using the
potential in eq.~\ref{eq:pot}\\
$\beta_{\ell}^{\pm}$: & resonance phase\\
$n_{\ell}^{\pm}$: & number of resonances\\
$\phi_{\ell}^{\pm}$: & phase relative to the hard sphere scattering \\
& for a  particular resonance\\
$\sigma_\ell$: & Coulomb phase of wave $\ell$\\
\end{tabular}

\vspace{3mm}
\noindent
The estimates of the widths of these states in $^{11}$N are based 
on the known widths of the analog states 
\begin{table}[htb]
 \caption{The resonances used in the simplified R-matrix treatment.}
 \begin{tabular}{r|ccc|cc}
\multicolumn{6}{l}{Potential model fit}\\
\hline 
& \multicolumn{3}{c|}{Potential parameters} & \multicolumn{2}{c}{Resonance}\\
  & $V_\ell$ & $r_0$=r$_{{\ell}s}$=r$_c$ & $a_0$=a$_{{\ell}s}$ & $E_r$ & $\Gamma_r$ \\
  & (MeV)    & (fm)           & (fm)       & (MeV) & (MeV)\\
\hline 
1/2$^+$ & -66.554 & 1.2 & 0.5 & 1.45  & 1.56 \\
1/2$^-$ & -41.286 & 1.2 & 0.6 & 2.13  & 0.89 \\
5/2$^+$ & -64.801 & 1.205 & 0.38 & 3.74 & 0.45 \\
\hline
\multicolumn{6}{l}{Resonances added in the R-matrix fit}\\
\hline
$^{*}$3/2$^{+}$ & \multicolumn{3}{l|}{} & 3.94 & 0.58 \\
3/2$^{-}$ & \multicolumn{3}{l|}{\hspace{1mm} Mirror of 2.69~MeV level $^{11}$Be} 
 & 4.33 & 0.27 \\
$^{*}($5/2$^{+})$ & \multicolumn{3}{l|}{\hspace{1mm} Mirror of 3.41~MeV
 level $^{11}$Be}  & 4.81 & 0.40 \\
$^{*}($7/2$^{-})$ & \multicolumn{3}{l|}{} & 5.4 & 0.25 \\
\hline \hline
\multicolumn{6}{l}{$^*$ \footnotesize The parameters for these states 
are only suggestions } \\ 
\multicolumn{6}{l}{\footnotesize which reproduce the observed cross section.} \\ 
\end{tabular}
\label{tab:rmatrix}
\end{table}
in $^{11}$Be. 
Inclusion of these states already accounts for the missing 
cross section up to 3.7~MeV, but the part above 3.7~MeV is still 
underestimated. 

\begin{figure}[htb]
  \centerline{\epsfig{file=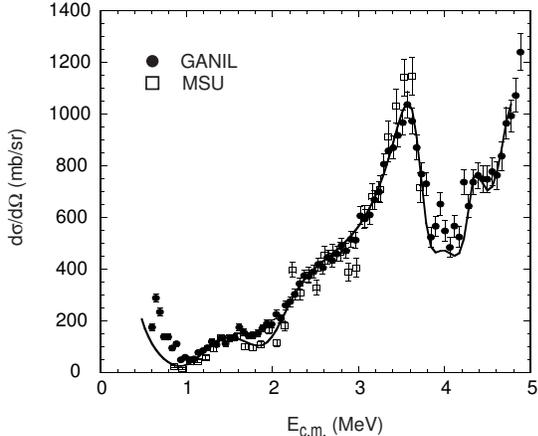, width=0.93\linewidth}}  
  \caption{The fit of the potential model with added resonances.
  The full drawn curve includes the two known resonances in $^{11}$Be 
  plus a broad 3/2$^{+}$ level around 4~MeV and a high-spin state at 5.4~MeV, 
  which gives a good
  description of the full excitation function up to 4.7~MeV. Especially
  it should be noted that the theoretical curve now better reproduces 
  the data at 1.6-2.0~MeV.}
  \label{fig:rmatrix}
\end{figure}

In particular, the energy region around 1.8~MeV is now
better reproduced, indicating that interference
of higher lying states indeed give the cross section that 
is not reproduced by the potential model in this region.
Inclusion of a 3/2$^{+}$ state at 3.94~MeV and a high-spin state
improves the description also at energies above 4~MeV, 
as is seen in Figure~\ref{fig:rmatrix}. 
The parameters for the potential and
included resonances are given in Table~\ref{tab:rmatrix}.

The conclusion drawn from comparing the results in Table~\ref{tab:para}
and Table~\ref{tab:rmatrix} is that
the positions and widths obtained using only the potential model 
are rather insensitive to the inclusion of higher states, which only
modifies the absolute cross section.
\\

Of the three resonances included in the calculations, only the one at
4.33~MeV is distinctly seen in the excitation function, see
Figure~\ref{fig:rmatrix}.
Its position corresponds to the 3/2$^{-}$ state at 2.69~MeV in $^{11}$Be within
150~keV, and the cross section supports a spin of 3/2 for this
resonance. The obtained width of 270~keV also agrees with the width of
the 2.69~MeV level in $^{11}$Be if decay by a $\ell$=1 proton is
assumed. We thus conclude that the narrow resonance at 4.33~MeV in
$^{11}$N is the mirror of the 2.69~MeV state in $^{11}$Be, both having
$I^{\pi}$=3/2$^{-}$.
The other resonances above 4~MeV are introduced in order to
reproduce the cross section at higher energies. The experimental data is
not sufficient to give conclusive determination of any parameters of
these states, but the existence of resonances above 4.4~MeV is 
necessary to reproduce the measured cross section.

\section{Discussion}\label{sec:discussion}
\subsection{The three lowest resonances in $^{11}$N.}
Table~\ref{tab:para} presents the parameters used in different fits of the
deduced excitation function in the $^{10}$C+p system. The fits were all
made under the assumption of three low-lying resonances. From these data
we conclude that the three lowest states in $^{11}$N have I$^\pi$= 1/2$^+$,
1/2$^-$ and 5/2$^+$. This is the first time all these states are identified in
one single experiment. However, there have been indications of them
in other reaction experiments.
In the pioneering work on $^{11}$N by Benenson et al.~\cite{ben74:9}, 
where the $^{14}$N($^3$He,$^6$He) reaction was studied, it was proposed that 
the resonance observed at 2.24~MeV ($\Gamma$=740 keV) was a 1/2$^{-}$ state.
This conclusion was based on the reaction mechanism in their experiment.
Our data confirms this result and both position and width are
within the experimental errors of the two experiments. 
The difference may probably be attributed to  different approaches 
in extracting the resonance parameters.
In a recent paper by Lepine-Szily et al.~\cite{lep98:80}, 
a state at 2.18~MeV was observed and interpreted as a 1/2$^{-}$ state, 
but with a width that was considerably narrower than in our work 
or that of Ref.~\cite{ben74:9}. 

The state at 1.27~MeV, which we interpret as a 1/2$^{+}$ state, 
could not be seen in the two experiments in Refs.~\cite{ben74:9,lep98:80},
as the selected
reactions quench the population of this state considerably. 
It could, however, be observed in an experiment performed at MSU where 
Azhari et al.~\cite{azh98:57} studied proton emission from $^{11}$N produced 
in a $^{9}$Be($^{12}$N,$^{11}$N) reaction. They found indications of a 
double peak at low energies and by fixing the upper part 
of it to the parameters from Ref.~\cite{ben74:9} they arrived at 
an excitation energy of 1.45~MeV.

The 5/2$^+$ (3.75~MeV) state was discussed in~\cite{axe96:54}.
The experiment presented in~\cite{lep98:80} showed a state at 3.63~MeV 
with a width about 400 keV. The position of the resonance is close to
ours but again the width is smaller in~\cite{lep98:80}.  

As well as for the 1/2$^{+}$ and 1/2$^{-}$ states, the
spin value for the 5/2$^+$ resonance does not leave any doubt 
that it is the mirror state of the 5/2$^+$ level at 1.78~MeV in $^{11}$Be 
($\Gamma$ = 100 keV). The potential model with the parameters used for $^{11}$N
and given in Table~\ref{tab:para}:c agrees very well for the width while the
excitation energy becomes 1.63~MeV. Still we consider this as an additional
support for our interpretation.

Fortune et al.~\cite{for95:51} have predicted the splitting between $0d_{5/2}$ and
$1s_{1/2}$ states in $^{11}$N from the systematics of this energy difference
for light nuclei, mainly assuming isobaric spin conservation.
Their results can therefore be considered as a direct extrapolation of experimental
data. The energy difference obtained in our work (${\Delta}E$=2.48~MeV) 
is close to the prediction of 2.3~MeV in~\cite{for95:51}. The energy 
difference between the $1s_{1/2}$ and $0p_{1/2}$ was calculated using the 
complex scaling method in~\cite{aoy98:57} and the value of 830 keV agrees 
with our data which gives 740 keV.

\subsection{Resonances above 4~MeV}
We interpret the structure around 4.3~MeV as due to a sharp resonance
in $^{11}$N, which is the mirror
state of the 2.69~MeV level in $^{11}$Be, Table~\ref{tab:rmatrix}. 
Several different experiments 
(see for example~\cite{liu90:42}) give the spin-parity for this
$^{11}$Be level as 3/2$^-$. 
The negative parity is well established from measurements of the
$^{11}$Li beta-decay~\cite{aoi97:616,mor97:627,Muk00:xxx}, and by 
measurements of the $^{9}$Be(t,p)$^{11}$Be
reaction \cite{liu90:42}. There is also good agreement between the Cohen-Kurath
prediction for the spectroscopic factor and the reduced single particle widths
of these mirror states. We found very good agreement between the widths if
the states undergo nucleon decay with $\ell\!=\!1$ ($j\!=\!3/2$). If the
states decay with 
orbital momentum, $\ell\!=\!2$ ($j=3/2$), the state in $^{11}$N will be at least twice
as broad, and in the case of $\ell\!=\!3$ ($j\!=\!5/2$) it would be at
least 3 times broader. 
Also, for $\ell\!=\!3$ the reduced single particle width will be too large, 
contradicting~\cite{zwi79:315}.
Using the simplified R-matrix approach, the position of the 3/2$^-$ level was
determined as 4.33~MeV.
The observed cross section for the population of this state is also in accordance
with a 3/2$^-$ assignment. The calculations further indicate that 
about one third of the
width of the 4.33~MeV state is due to the proton decay to the
first excited state in $^{10}$C.
Even a small branch of this decay results in a large reduced
width. This indicates a large coupling of 3/2$^-$ state to the first excited
state of the core, as was recently predicted by Descouvemont~\cite{des97:615}.
In~\cite{axe96:54} we proposed a different structure for
the 3/2$^-$ state (two particles in the $1s$ state) because preliminary treatment
resulted in a too small width (70~keV) for the state. 

In the present experiment there is an experimental cutoff at 5.4~MeV (see
Figure~\ref{fig:nofit}) and the excitation function increases towards 
this high-energy end. This is most likely due to 
higher-lying states but we cannot make any assignments for 
them based on our data.
However the authors \cite{azh98:57} had to introduce a broad 
($\Gamma$= 500-1000 keV)
state in the energy region around 4.6~MeV
to explain the spectrum from $^{11}$N decay. They
proposed the broad state to be a 3/2$^-$ state. Our data show that the
3/2$^-$ state in $^{11}$N 
is rather narrow, and therefore another state has to be assumed 
in order to explain
the data in~\cite{azh98:57}. This is also a justification for the inclusion
of the 3/2$^+$ resonance is our R-matrix fit.

Various theoretical calculations (for recent references
see~\cite{for99:461}) have attempted 
to reproduce the level sequence in $^{11}$Be. 
Most models emphasize the role of coupling between the
valence neutron and the first excited 2$^+$ state 
in $^{10}$Be in generating the
parity inversion. It is well known that model assumptions influence
the wave functions more than their energy eigenvalues
and thus models giving the correct level sequence predict 
very different core excitation
admixtures. For the ground state in $^{11}$Be, the admixtures
given by theoretical calculations vary from
7\%~\cite{des97:615,vin95:592} 
to 75\%~\cite{rag90}. Theoretical results are frequently 
compared to spectroscopic factors obtained from 
nucleon transfer reactions. The single-particle
spectroscopic factors for $^{11}$Be 
have been obtained from $^{10}$Be(d,p) reactions~\cite{zwi79:315}. 
Even if the theory of stripping reactions
is very well developed, many parameters are involved 
in the extraction of these results from the data.
Evaluating single-particle nucleon widths using a potential
model involves fewer parameters. 
For the lowest states of $^{11}$N we obtained the
reduced widths given in Table~\ref{tab:sf}. For the s$_{1/2}$ state
we have a reduced width of~$\approx$1 which, taking the 15\% 
experimental error in the width into account, indicates that
no large core-excitation admixtures are needed to describe
the ground states of $^{11}$N and $^{11}$Be.

\section{Summary}\label{sec:summary}
The excitation function in the $^{10}$C+p system has been studied
using elastic resonance scattering. The low-energy part was analyzed in 
a potential model while the high-energy part was described in a
simplified R-matrix approach. The ground state and the first two excited states in
the unbound nucleus $^{11}$N was found to have the spin-parity sequence of
1/2$^{+}$, 1/2$^{-}$ and 5/2$^{+}$ which is identical to that found 
in its mirror partner $^{11}$Be. A narrow 3/2$^-$ state at 4.33~MeV
was identified as the mirror state of the 2.69~MeV state in
$^{11}$Be. The energies and widths of the observed states are listed
in Table~\ref{tab:comp}. The agreement among experiments as well as between
our results and theoretical calculations 
\onecolumn
\twocolumn[\hsize\textwidth\columnwidth\hsize\csname@twocolumnfalse\endcsname 
\begin{table}
\caption{A summary of all experimental and theoretical results on
  $^{11}$N. $E_r$ and $\Gamma_r$ denote the resonance energy and the
  width of the resonance, respectively. The error bars in this work
include the systematic errors (25 keV in the c.m. frame) as well as the
spread in results obtained for different potential parameters that all
fit the data.}
\begin{center}  
\begin{tabular}{l|cc|cc|cc}
 & \multicolumn{2}{c|}{1/2$^{+}$} & \multicolumn{2}{c|}{1/2$^{-}$} 
&\multicolumn{2}{c}{5/2$^{+}$}  \\
Ref.& $E_r$ & $\Gamma_r$ & $E_r$ & $\Gamma_r$ & $E_r$ &  $\Gamma_r$ \\
\hline
\multicolumn{7}{l}{Experimental papers}   \\
\hline
\small{This work} & $1.27^{+0.18}_{-0.05}$ & $1.44\pm0.2$ &$2.01^{+0.15}_{-0.05}$  
& $0.84\pm0.2$ & $3.75\pm0.05$ & $0.60\pm0.05$  \\
\hline
\cite{ben74:9}& - & - & $2.24\pm0.1$  & $0.74\pm0.1$   & - & - \\
\cite{lep98:80}& - & - & $2.18\pm0.05$ & $0.44\pm0.08$ & $3.63\pm0.05$ &
$0.40\pm0.08$ \\
\cite{azh98:57}$^a$& $1.45\pm0.40$ & $>0.4$ & $2.24\pm0.1$ & $0.74\pm0.1$& - & -  \\
\hline
\multicolumn{7}{l}{Theoretical papers}   \\
\hline
\cite{she95}$^a$& 1.54 & 0.62 & $2.24\pm0.1$ & $0.74\pm0.1$ & 3.74 & 0.3  \\
\cite{for95:51}& $1.60\pm0.22$ & $2.1^{+1.0}_{-0.7}$ & 2.49 & 1.45 & 3.90 & 0.88  \\
\cite{bar96:53}$^b$& 1.4 & 1.31 & 2.21 & 0.91 & 3.88 & 0.72  \\
\cite{des97:615}& 1.1 & 0.9 & 1.6 & 0.3 & 3.8 & 0.6  \\
\cite{gre97:56}& 1.2 & 1.2 & 2.1 & 1.0 & 3.7 & 1.0  \\
\hline\hline
\multicolumn{7}{l}{\footnotesize $^{a}$For the $1/2^-$ state the
  results from~\cite{ben74:9}.}\\
\multicolumn{7}{l}{\footnotesize $^{b}$The results obtained with
  $r_0=1.45$~fm is presented.}\\
\end{tabular}
\end{center}
\label{tab:comp}
\end{table}
]
\noindent
are very satisfactory.

The quasi-stationary character of $^{11}$N states was used to evaluate 
the reduced single-particle widths for the identified states. This result
indicates small coupling between the valence nucleon in the ground state
$^{11}$N and the first excited 2$^+$ state in $^{10}$C, and the same
conclusion should be valid for $^{11}$Be.
 
The experimental technique to use elastic-resonance scattering with
radioactive beams has proven to be a very efficient tool for 
investigations beyond the dripline.

\section{Acknowledgments}
The authors thank Prof.~M.~Zhukov and Prof.~F.~Barker for valuable
discussions. 
The work was partially supported by the National Science
Foundation under grant PHY95-28844. 
The work was also partly supported by a grant from RFBR.
S.~B. acknowledges the support of
the REU program under grant PHY94-24140.
\vspace{-5mm}\\ 

\end{document}